\documentclass[10pt,twocolumn,floatfix]{revtex4-1}
\usepackage{graphicx}
\usepackage{epstopdf}
\usepackage{amssymb}
\usepackage{mathrsfs}
\usepackage{amsmath}
\usepackage{color}
\usepackage{latexsym}
\usepackage{amsfonts}
\usepackage{hyperref}
\usepackage{wasysym}
\usepackage{dcolumn}
\usepackage{bm}
\usepackage{hyperref}
\usepackage{soul}
\usepackage[caption=false]{subfig}
\usepackage{url}

\usepackage{natbib}

\newcommand{\mr}{\mathrm}

\begin{document}

\selectfont

\title{Multi-scale approach for self-Assembly and protein folding}

\author{Oriol Vilanova$^1$}
\author{Valentino Bianco$^2$}
\author{Giancarlo~Franzese$^1$}

\email{gfranzese@ub.edu}

\affiliation{ $^1$Secci\'o de
 F\'isica Estad\'istica i Interdisciplin\`aria--Departament de F\'isica
 de la Mat\`eria Condensada, Facultat de F\'isica \& Institute of Nanoscience and Nanotechnology (IN2UB),
  Universitat de Barcelona, Mart\'i i Franqu\`es 1, 08028 Barcelona, Spain}

\affiliation{$^2$ Computational
  Physics Group, Faculty of Physics, Universit\"at Wien, Sensengasse 8/10, 1090 Vienna,
  Austria}

\date{\today}

\definecolor{corr14okt}{rgb}{0,0,1} 
\definecolor{moved}{rgb}{0,0,0}

\begin{abstract} 
  We develop a multi-scale approach to simulate hydrated nanobio
  systems under realistic conditions (e.g., nanoparticles and protein 
solutions at physiological conditions over time-scales up to
hours). We combine atomistic simulations of water at bio-interfaces
(e.g., proteins or membranes) and nano-interfaces (e.g., nanoparticles
or graphene sheets) and coarse-grain models of hydration water for
protein folding and protein design. We study protein self-assembly
and crystallization, in bulk or under confinement, and the kinetics
of protein adsorption onto nanoparticles, verifying our predictions in
collaboration with several experimental groups. We try to find answers
for fundamental questions (Why water is so important for life? Which
properties make water unique for biological processes?) and
applications (Can we design better drugs? Can we limit
protein-aggregations causing Alzheimer? How to implement
nanotheranostic?). Here we focus only on the two larger scales of our
approach: (i) The coarse-grain description of hydrated proteins and
protein folding at sub-nanometric length-scale and
milliseconds-to-seconds time-scales, and (ii) the coarse-grain
modeling of protein self-assembly on nanoparticles at
10-to-100 nm length-scale and seconds-to-hours time-scales.
\end{abstract}

\maketitle

\section{Introduction}
Self-assembly, driven by non-covalent interactions like van der Walls
and hydrogen bonds, fulfills a crucial role in the supramulecular
organization and assembling of the biological matter. Living being are
complex organisms where matter is self-organized on different
length scales in a kind of biological network. A primary role is played
by the proteins that  control the majority of chemical processes in
the cell. Proteins are synthesized as long polymer chains composed by
hundreds of monomers, taken from 20 different amino acids.  
Among the huge amount of possible amino acid sequences, nature has
selected those that are able to fold into specific functionalized
structures, known as native protein structures. The relation between
the sequence code and the native structure and the way that proteins
fold represents a significant example of biological self-assembly. 

A crucial aspect of the protein folding that involves the interplay
with solvent is related to the hydrophobic or hydrophilic nature of
the amino acids.  
The protein folding mechanism is dominated by the dynamics of water
that drives the collapse of the hydrophobic protein core and stabilizes
the tertiary protein structure \cite{LevyPNAS2004, Levy2006,
  Raschke2006}. However, many proteins exhibit a limited range of
temperatures $T$ and pressures $P$ where they are able to maintain the
native structure \cite{Zipp1973, privalov, HummerPNAS1998,
  MeersmanHighPressRes2000, Lassalle2000, Smeller2002, Herberhold2002,
  Lesch2002, RavindraChemPhysChem2003, MeersmanChemSocRev2006, PastoreJACS2007,
  WiedersichPNAS2008, Maeno2009, Somkuti2013, Somkuti2013a,
  NucciPNAS2014}. Beyond those $T$-- and $P$--ranges a protein
unfolds, with a consequent loss of its tertiary structure and
functionality.

At high $T$ protein unfolding is due to the thermal fluctuations that disrupt the protein structure. Open protein conformations increase the entropy $S$ minimizing the global Gibbs free energy $G\equiv H-TS$, where $H$ is the total enthalpy.
By decreasing $T$ proteins can crystallize but, if the nucleation of water is avoided some proteins denaturate \cite{privalov, Griko1988, RavindraChemPhysChem2003, goossens, nash, nash2, MeersmanHighPressRes2000, PastoreJACS2007}. Usually such phenomena are observed below the melting line of water, although in some cases cold denaturation occurs above the $0^\circ$, as in the case of the yeast frataxin \cite{PastoreJACS2007}.

Cold- and $P$-denaturation of proteins have been related to the equilibrium properties of
hydration water \cite{delosriosPRE2000, marquesPRL2003,
  PatelBPJ2007, Athawale2007, NettelsPNAS2009, Best2010,
  Jamadagni2010, Badasyan2011, matysiakJPCB2012, BiancoJBioPhys2012,
  BiancoPRL2015}. 
However, the interpretation of this mechanism is still largely debated
\cite{PaschekPRL2004, Paschek2005, Sumi2011, Coluzza2011, Dias2012,
  DasJPCB2012, SarmaCP2012, FranzeseFood2013, Abeln2014, Yang2014,
  RochePNAS2012, Nisius2012, 2015arXiv151106590V}. 

Protein denaturation is observed also upon pressurization \cite{Zipp1973, HummerPNAS1998, PaschekPRL2004, MeersmanChemSocRev2006, NucciPNAS2014}. A possible explanation of the high-$P$ is the loss of  internal cavities, sometimes presents in the folded states of proteins \cite{RochePNAS2012}. Denaturation at negative $P$ has been experimentally observed \cite{Larios2010} and simulated \cite{Larios2010, HatchJPCB2014, BiancoPRL2015} recently.
Pressure denaturation is usually observed at 100 MPa $\lesssim P \lesssim$ 600 MPa, and rarely at higher $P$ unless the tertiary structure is engineered with stronger covalent bonds \cite{Lesch2002}.

\section{Hawley theory}
\begin{figure}
 \includegraphics[width=\linewidth]{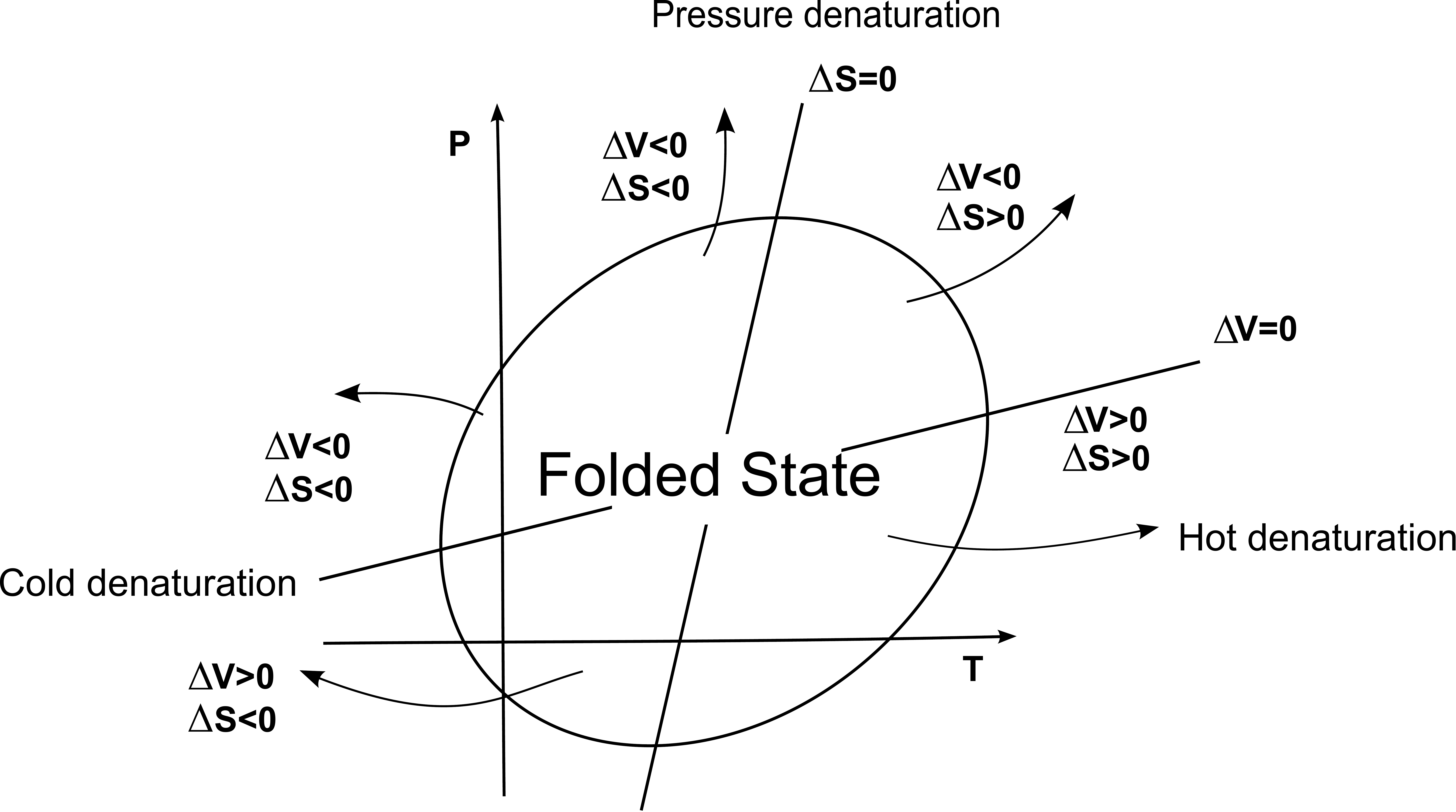}
 \caption{Stability region of proteins according to the Hawley theory \cite{hawley}. A protein is stable in its native state for temperature and pressures inside the elliptic curve. The denaturation occurs with a positive or negative variation of the total (protein plus the solvent) volume $\Delta V$ and total entropy $\Delta S$ according to the figure. Staight lines join the point where the $f \longrightarrow u$ transition is isoentropic or isochoric, respectively where the slope of the coexistence line is vanish or infinite. 
 Adapted from \cite{BiancoJBioPhys2012}.}
 \label{hawley_theory}
\end{figure}

In 1971, Hawley proposed a theory \cite{hawley} explaining the close stability region (SR) for proteins in the $T$--$P$ plane (Fig \ref{hawley_theory}). The SR represents the region where a protein folds into its native conformation. Such a simple two state theory is based on the assumption that the folding $(f)$ unfolding $(u)$ transition is a first order phase transition and that equilibrium thermodynamics holds during the denaturation, neglecting all the details about the protein structure. Following \cite{hawley} we can express the free energy difference $\Delta G\equiv G_f - G_u$ between the free energy of the folded ($G_f$) and the unfolded ($G_u$) states as a quadratic function of $T$ and $P$
\begin{equation}\label{taylor}
\begin{array}{cc}
\Delta G(P,T)=\dfrac{\Delta\beta}{2}(P-P_{0})^{2} + 2\Delta\alpha(P-P_{0})(T-T_{0}) + \\ \\ -\dfrac{\Delta C_{P}}{2T_{0}}\left(T-T_{0}\right)^{2}+ \Delta V_{0}(P-P_{0}) -\Delta S_{0}(T-T_{0})+ \Delta G_{0}
\end{array}
\end{equation}
where $T_0$ and $P_0$ are the temperature and pressure of the reference state point (ambient conditions); $\Delta V$ and $\Delta S$ are the volume and entropy variation upon unfolding respectively;  $\alpha \equiv (\partial V/\partial T) = -(\partial S/\partial P)_T$ is the thermal expansivity factor, related to the isobaric thermal expansion coefficient $\alpha_P$ by $\alpha_P = \alpha/V$; $C_P \equiv T(\partial S/\partial T)_P$ is the isobaric heat capacity; $\beta \equiv ≡ (\partial V/\partial P)_T$ is the isothermal compressibility factor related to the isothermal compressibility $K_T$ by the relation $K_T = −(\beta/V)$ and $\Delta G_0$ is an integration constant. 
Eq. (\ref{taylor}) represents an ellipsis given the constrain 
\begin{equation}
 \Delta \alpha^2 > \Delta C_P\Delta \beta / T_0
\end{equation}
which is guaranteed by the different sign of $\Delta C_P$ and $\Delta \beta$ as reported by Hawley \cite{hawley}. Although the calculation of Hawley is based on a taylor expansion of the free energy variation truncated to  second order. Adding more terms in the Eq. (\ref{taylor}) results in minor corrections that do not affect the close elliptic-like coexistence  curve.
All in all, the Hawley model is a phenomenological theory that makes strong assumptions on the $f\longrightarrow u$ process \cite{Meersman2006}. Nevertheless, its ability to describe all the denaturation mechanisms actually observed in experiments makes it a good test for  models of protein unfolding.

\begin{figure*}
\includegraphics[scale=1]{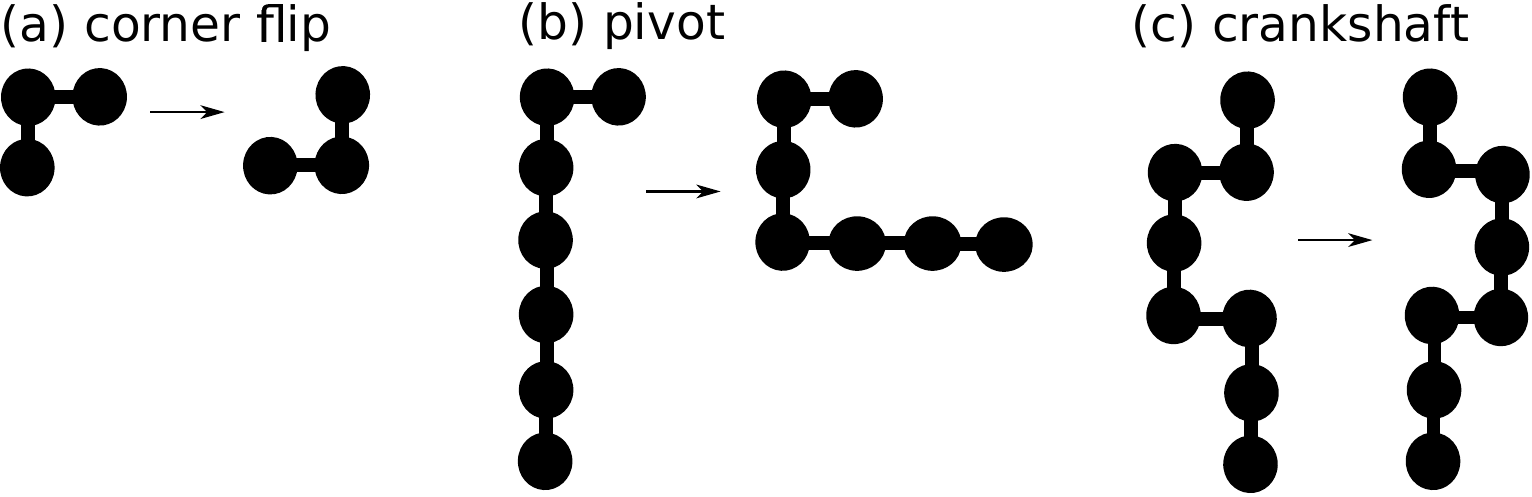}
\caption{ Possible protein move. (a) In the {\it corner flip} move a monomer in a corner configuration can jump to the opposite corner. (b) In the {\it pivot} move we randomly choose one monomer, acting as a pivot, and rotate the shorter side chain, with respect to it. (c) The {\it crankshaft} move consists in choosing at random an axes passing through two monomers, and rotating the included residues. }
\label{protein_move}
\end{figure*}

\section{A coarse grain model for solvated protein}
 
\subsection{Bulk water model}
We present a coarse-grain model for protein water interactions based
on a lattice representation of the protein, embedded in explit
water. This water model adopted is ``many-body''
\cite{StokelyPNAS2010, strekalovaPRL2011, FranzeseFood2011,
  MazzaPNAS2011, FranzeseFood2013, BiancoProceedings2013,
  delosSantos2011, BiancoSR2014, BiancoPRL2015, 2016arXiv161000419C}.
  
The coarse-grain representation of the many-body interactions is based on a discretization of the available molecular volume  $V$ into a fixed number $N_0$ of cells, each with volume $v\equiv V/N_0\geq v_0$, where $v_0$ is the water excluded volume. 
Each cell accommodates at most one molecule with the average O--O
distance between next neighbor water molecules given by $r=v^{1/3}$. 
To each cell we associate a variable $n_{i}=1$ if the cell $i$ is
occupied by a water molecule and it has $v_0/v> 0.5$, and $n_{i}=0$
otherwise. Hence, $n_{i}$ is a discretized density field replacing the
water translational degrees of freedom.  
The Hamiltonian of bulk water is
\begin{equation}
\mathscr{H}\equiv \sum_{ij} U(r_{ij}) -J N_{\rm HB}^{\rm (b)}-J_\sigma N_{\rm  coop}.
\label{bulk}
\end{equation}
The first term  accounts for the van der Waals interaction and is modeled with a Lennad-Jones potential 
\begin{equation}
  \sum_{ij} U(r_{ij})\equiv 4\epsilon\sum_{ij}\left[\left(\frac{r_0}{r_{ij}}\right)^{12}-\left(\frac{r_0}{r_{ij}}^6\right)\right]
\end{equation}
where the sum runs over all the water molecules $i$ and $j$ at O--O distance $r_{ij}$ and $\epsilon \equiv 5.8$ kJ/mol. We assume $U(r)\equiv \infty$ for $r<r_0\equiv v_0^{1/3}= 2.9 $ \AA{} that is the water molecule hard core, (water van der Waals
diameter). Moreover, we apply a cutoff to the potential for $r>r_c \equiv 6 r_0$. 

The second term represents the directional and covalent components of
the hydrogen bond (HB), where 
\begin{equation}
 N_{\rm HB}^{\rm (b)}\equiv\sum_{\langle ij \rangle}n_i n_j\delta_{\sigma_{ij},\sigma_{ji}}
\end{equation}
is the number of bulk HBs and the sum runs over the neighboring
cells. $\sigma_{ij}=1, \dots, q$ is the bonding index of molecule $i$
with respect to the neighbor molecule $j$. $\delta_{ab}=1$ if $a=b$, 0
otherwise. Each water molecule can form up to four HBs. Each HB is
stable if the hydrogen atom H is in a range of $[-30^\circ;30^\circ]$
with respect to the O--O axes. Hence, only $1/6$ of the entire range
of values $[0,360^\circ]$ for  the ${\widehat{\rm OOH}}$ angle is
associated to a bonded state, leading to the choice $q=6$ to account
correctly for the entropy variation due to  HB formation and
breaking. 
For each HB the energy decreases an amount $-J$, where $J/4\epsilon= 0.3$. According to Ref. \cite{StokelyPNAS2010}, a good choice for the parameters is $\epsilon =5.5$ kJ/mol,  $J/4\epsilon=0.5$ and $J_\sigma/4\epsilon = 0.05$. For such a choice, the average HB energy is $\sim23$ kJ/mol. Here, to account for the ions in solution that are always present in the cellular environment where natural proteins are embedded, we decrease the ratio $J/J_\sigma$, that modify the the bulk phase diagram in a qualitative way similar to that induced by ions \cite{corradiniJPCB2011}. For such a choice we find that the HB energy is $\sim 20$ kJ/mol.

The third term in the Hamiltonian represents the cooperative interaction between the HBs. Such an effect is due to the quantum many-body interaction \cite{HernandezJACS2005}: the formation of a new HB affects the electron distribution around the molecule favoring the formation of the following HB in a local tetrahedral structure \cite{SoperPRL2000}.
To mimic the cooperativity of the HBs we introduce an effective interaction between the bonding indexes of a molecule
\begin{equation}
 N_{\rm coop}\equiv \sum_i n_i\sum_{(l,k)_i}\delta_{\sigma_{ik},\sigma_{il}}
\end{equation}
where $(l,k)_i$ indicates each of the six different pairs of the four indices $\sigma_{ij}$ of a molecule $i$. The choice  $J_\sigma/4\epsilon\equiv 0.05\ll J$  guarantees the asymmetry between the two HB terms.

The formation of HBs leads to an open network of molecules, giving rise to a lower density state. We include this effect into the model assuming that, for each HB formed, the volume $V$ increases by $v_{\rm HB}^{\rm (b)}/v_0=0.5$, corresponding to the average volume increase between
high-density ices VI and VIII and low-density (tetrahedral) ice Ih. We assume that the HBs do not affect the distance $r$ between first neighbour molecules, consistent with experiments \cite{SoperPRL2000}. Hence, the HB formation does not affect the $U(r)$ term.

The total bulk volume $V^{\rm (b)}$ is 
\begin{equation}
 V^{\rm (b)}\equiv Nv_0+N_{\rm HB}^{\rm (b)}v_{\rm HB}^{\rm (b)}.
\end{equation}

\subsection{Modeling protein-water interplay}
The protein is modeled as a hydrophobic self-avoiding lattice polymer and it is embedded into the cell partition of the system. Despite its simplicity, lattice protein models are still widely used in the contest of protein folding \cite{lauMacromol1989, caldarelliJBioPhys2001, marquesPRL2003, PatelBPJ2007, matysiakJPCB2012, BiancoPRL2015} because of their versatility and the possibility to better understand many mechanisms of the protein dynamics. Each protein residue (polymer bead) occupies one cell, without affecting its volume. In the present study, we do not consider the presence of cavities into the protein structure.

To simplify the discussion, we assume that no residue-residue interactions occur and that the residue-water interaction vanishes. 
This implies that the protein has several ground states, all with the same maximum number $n_{\rm max}$ of residue-residue contacts.  
Our results holds also when such interactions are restored \cite{BiancoPRL2015}. Here we adopt the symbol $\Phi$ to refer to hydrophobic residues.
The protein interface affects the water-water properties in the hydration shell, here we define it as the layer of first neighbor water molecules in contact with the protein. 
There are many numerical and experimental evidences supporting the hypothesis that water-water HBs in the hydration shell are more stable and more correlated with respect to bulk HBs \cite{DiasPRL2008, PetersenJCP2009, SarupriaPRL2009, TarasevichCollJ2011, DavisNat2012}. 
We account for this by replacing $J$ of Eq.~(\ref{bulk}) with $J_{\Phi}>J$ for water-water HBs at the $\Phi$ interface.
This choice, according to Muller \cite{muller1990}, ensures the water enthalpy compensation upon  cold-denaturation
\cite{BiancoJBioPhys2012}. 

In addition to the stronger/stabler water-water HBs in the $\Phi$
shell, we incorporate into the model also the larger density
fluctuations at the $\Phi$ interface with respect to the bulk,
observed in water hydrating $\Phi$ solutes \cite{SarupriaPRL2009,
  DasJPCB2012}. Such an increase of density fluctuations results in a
$\Phi$ hydration shell that, at ambient pressure, is more compressible
than bulk water. Although it is still matter of debate whether the average
density of water at the $\Phi$ interface is larger or smaller with
respect to the average bulk water density \cite{Lum:1999kx,
  Schwendel2003, jensenPRL2003, DoshiPNAS2005, Godawat2009}, there is
evidence showing that such density fluctuations reduce upon
pressurization \cite{SarupriaPRL2009, DasJPCB2012, GhoshJACS2001,
  DiasJPCB2014}. 
Hence,  if we attribute this $P$-dependence of the $\Phi$ shell density to the interfacial HB properties, we can assume that the average volume associated to HBs formed in the $\Phi$ shell is
\begin{equation}\label{vsurf}
 v_{\rm HB}^{(\Phi)}/v_{\rm HB,0}^{(\Phi)}\equiv 1- k_1P
\end{equation} 
where $v_{\rm HB,0}^{(\Phi)}$ is the volume change associated to the
HB formation in the $\Phi$ hydration shell at $P=0$ and $k_1$ is a
positive factor. According to Ref. \cite{BiancoPRL2015}, we could add
other polynomial terms to the Eq. (\ref{vsurf}), althoug such terms
would not affect our results as long as $P<1/k_1$ \cite{2017arXiv170403370B}.

The total volume $V$, including the contributions coming from the HBs formed in the $\Phi$ shell is
\begin{equation}
 V\equiv Nv_0+N_{\rm HB}^{\rm (b)}v_{\rm HB}^{\rm (b)} +N_{\rm HB}^{(\Phi)}v_{\rm HB}^{ (\Phi)}.
\end{equation}
where $N_{\rm HB}^{(\Phi)}$ is the number of HBs in the $\Phi$ shell.

In the following we fix 
$k_1= 1 v_0/4\epsilon$, $v_{\rm HB,0}^{(\Phi)}/v_0=v_{\rm HB}^{\rm (b)}/v_0=0.5$ and $J_{\Phi}/J=1.83$. Our findings are robust with respect to a change of parameters.

\begin{figure}
\includegraphics[scale=0.4,bb=0 10 600 450,clip=true]{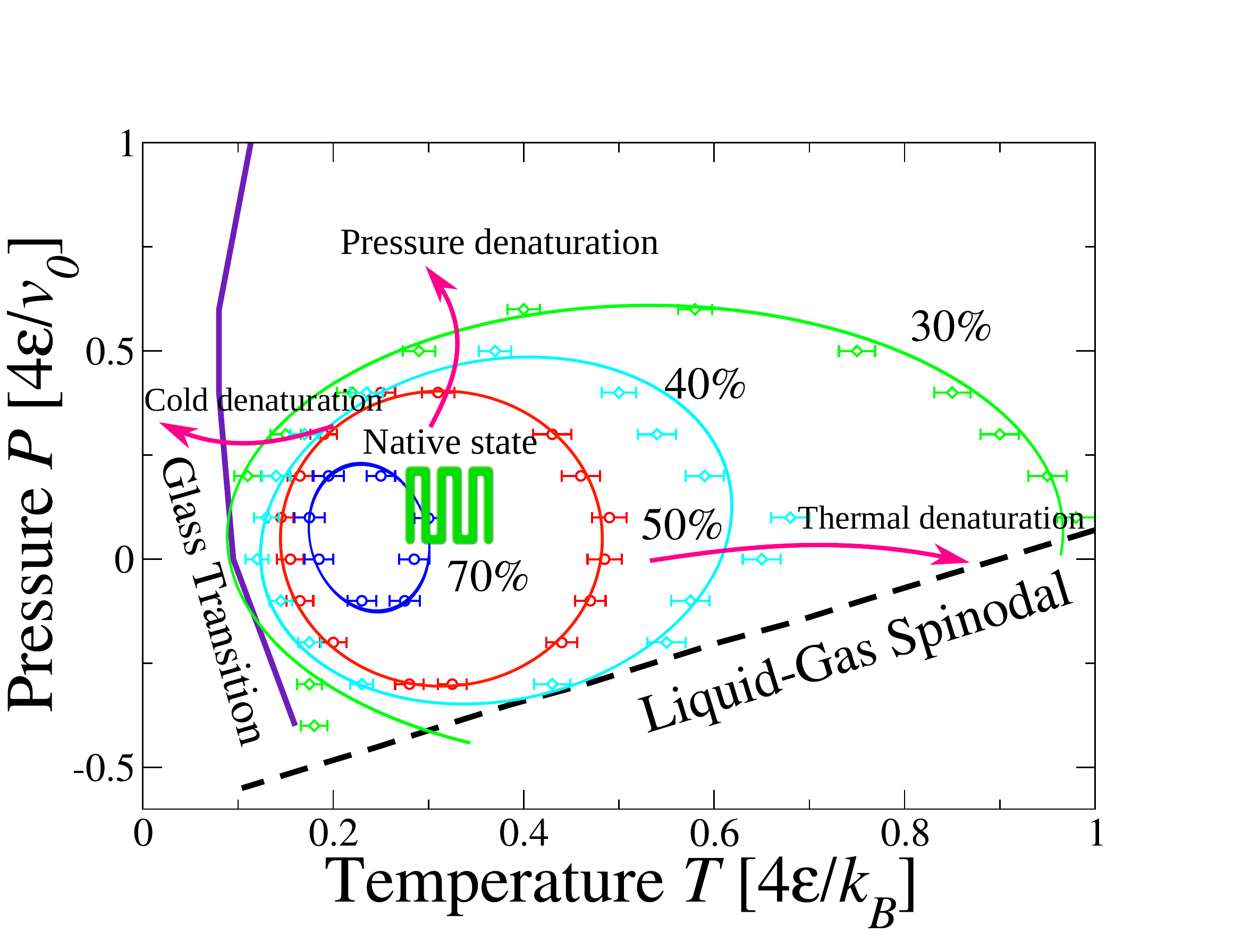}
\caption{ $P-T$ stability region of the protein, calculated with Monte Carlo simulations.
  The symbols   mark the state points where the protein has the same     average residue-residue contact's
  number $n_{\rm rr}/n_{\rm  max}=30\%$, 40\%, 50\% and 70\%, corresponding to different percentage of compactness.
Elliptic lines are guides for the eyes. The ``glass transition'' line
defines the temperatures below which the system does not 
  equilibrate. 
The spinodal line marks the stability limit of the liquid
phase at high $P$ with respect to the gas at low $P$; 
$k_B$ is
the Boltzmann constant.} 
\label{SR}
\end{figure}

\subsection{Simulation details}

We study proteins with 30 residues using Monte Carlo simulations in the isobaric-isothermal ensemble, i.e. at constant $P$, constant $T$ and constant number of particles. All the simulations start with a protein in a completely folded conformation. The water bonding indexes $\sigma$ are equilibrated using a cluster algorithm. The protein is equilibrated using corner flips, pivot and crunkshuft moves \cite{Frenkel2002b} (Fig \ref{protein_move}).
Along the simulation we calculate the average number of residue-residue contact points to estimate the protein compactness.
For each state point we sample $\sim 10^4$ independent protein conformations.

\subsection{Results}

We assume that the protein is folded if the average number of residue-residue contacts is
$n_{\rm rr}\geq 50\% ~n_{\rm  max}$.
In Fig. \ref{SR} we show the calculated SR for the protein, consistent
with the Hawley theory \cite{hawley,Smeller2002}. We observe that the
SR has an elliptic shape that is preserved independently of the
compactness we adopt as the reference for the folded state, underlying
that the folded$\longrightarrow$unfolded transition is a continuous
process. Proteins undergo heat-, cold-, and $P$-unfolding.  
The folded protein minimizes the number of hydrated $\Phi$ residues, reducing the energy cost of the interface, as expected.

Upon increasing $T$ at constant $P$, we observe that the model reproduces the expected 
{\em  entropy-driven} unfolding. The entropy $S$ increases both 
for the opening of the protein and for the larger decrease of
 water-water HBs.

Decreasing  $T$ at constant $P$ leads to open protein conformations
that minimize the Gibbs free energy. 
The difference in energy gain between bulk HBs  and HBs at the $\Phi$
interface results in a competing mechanism. A bulk water molecule can
form up to 4 HBs, while the water molecules at the $\Phi$ interface can
form up to 3 HBs, although stronger. Hence, reducing the exposed
protein surface maximizes the possible number of bulk HBs, while
increasing the hydrated protein surface maximizes the interfacial
HBs. 
At low $T$ the number of HBs in the $\Phi$ shell saturates and the
only way for the system to further minimize 
the internal energy is by increasing $N_{\rm  HB}^{(\Phi)}$, i.e. by
unfolding the protein. Hence, the cold denaturation is an {\em energy-driven}
process.

Upon an isothermal increase of $P$ the protein denaturates. 
We find that this change is associated to a decrease of $N_{\rm
  HB}^{\rm (b)}$ and an increase of 
$N_{\rm  HB}^{\rm (\Phi)}$ leading to a net decrease of $V$ at
high $P$, as a consequence of the compressible $\Phi$ shell, Eqs.~(\ref{vsurf}).
At high $P$, the $PV$ term of the Gibbs free energy dominates the $f
\longrightarrow u$ process. Hence, the water contribution to  
the high-$P$ denaturation induces a  {\em density-driven} process,
resulting in a denser $\Phi$ shell with respect to the bulk.

Finally, lowering $P$ toward negative values results in a negative 
 contribution  $(Pv_{\rm  HB}^{(\Phi)}-J_\Phi)N_{\rm HB}^{(\Phi)}$,
leading to a decrease in
enthalpy  when the protein opens up and $N_{\rm HB}^{(\Phi)}$
increases. Therefore we find that 
at negative $P$ the denaturation process 
is {\em enthalpy-driven}.

\begin{figure}
\vspace{-0.9 cm}
\includegraphics[scale=0.4]{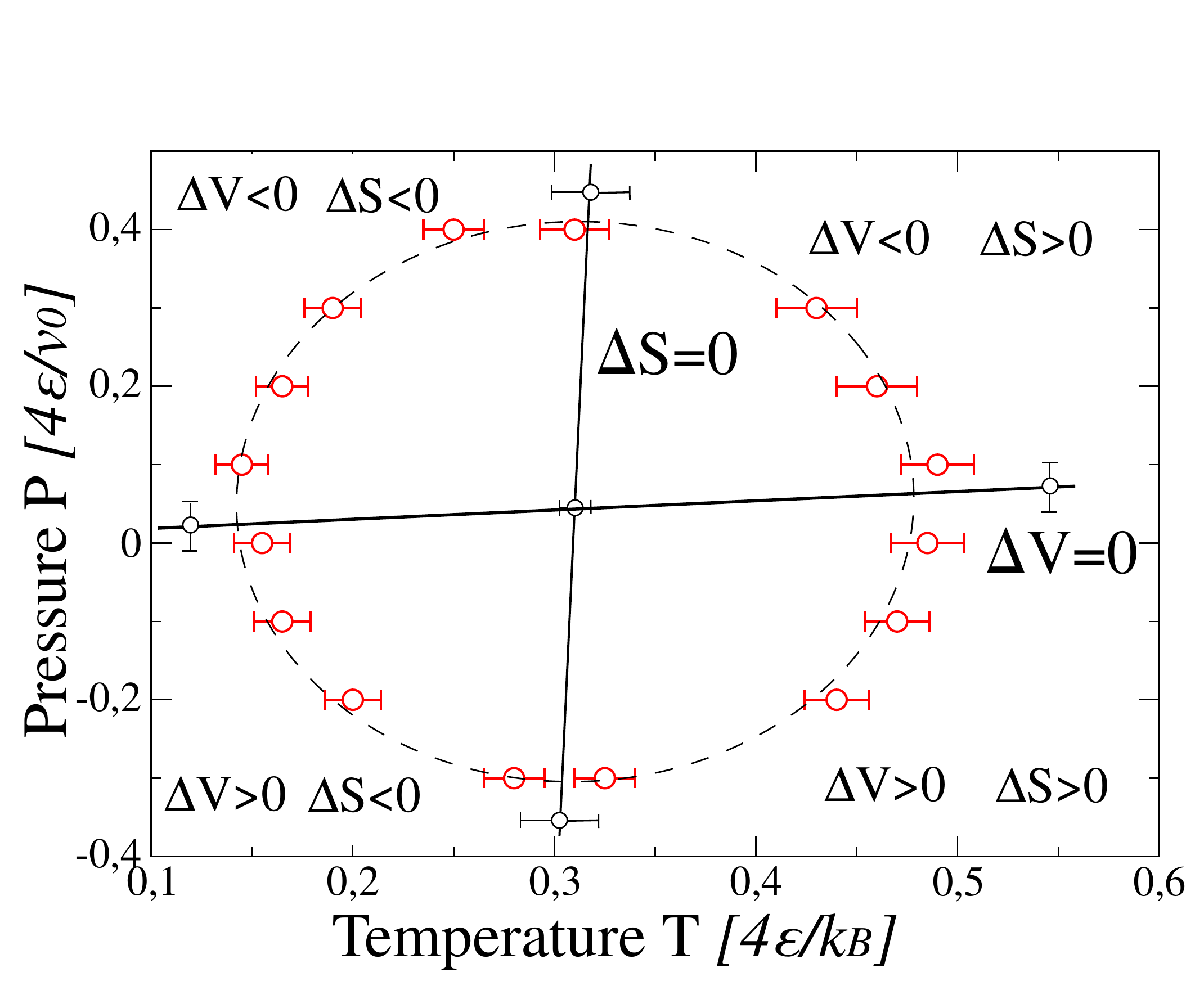}
\caption{Volume change $\Delta V$ and entropy change $\Delta S$  for the
  $f \longrightarrow u$ process in the $T-P$ plane.
Solid lines connect state points with isochoric $\Delta V =0$ and isoentropic $\Delta S =0$ denaturation.  Red points mark the SR, adopted from Fig. (\ref{SR}). 
The loci $\Delta V=0$ and $\Delta S=0$ have a positive slope and intersect the SR at
the turning points with  $dT/dP|_{\rm SR}=0$ and $dT/dP|_{\rm SR}=\infty$ respectively.}
\label{deltaV_S}
\end{figure}

By varying the parameters $v_{\rm HB}^{(\Phi)}$ and $J_{\Phi}$
we find that the first is relevant for the $P$-denaturation, as we expected
because it dominates the volume contribution to the Gibbs free energy, while the second affects the
stability range in $T$. Both effects combine in a non-trivial way to regulate
the SR, shifting, shrinking and dilating the SR, although the elliptic
shape is preserved \cite{2017arXiv170403370B}.

We can estimate also the entropy change and volume change, respectively indicated with $\Delta S$ and $\Delta V$, during the $f \longleftrightarrow u$ process. 
First, we calculate the average volume of the unfolded--completely stretched--protein $V_u$ and of the folded--maximum compactness--protein $V_f$ in a wide range of $T$ and $P$, equilibrating water  around the fixed protein conformations. From the difference $\Delta V \equiv V_f - V_u$ we calculate $\Delta S$ using the Clapeyron relation $dP/dT = \Delta S/ \Delta V$ applied to
the SR \cite{hawley}. The Clapeyron equation holds, in principle, only along first order phase transitions. Though the $f \longleftrightarrow u$ process is not necessary a phase transitions, the calculation of $\Delta S$ and $\Delta V$ represents a good model test to compare our model results with the Hawley's theory \cite{hawley}.
In Fig. \ref{deltaV_S} we show that our findings match with the theoretical predictions: $T$-denaturation is accompanied by a positive entropy variation $\Delta S>0$ at high $T$  and an entropic penalty $\Delta S<0$ at low $T$;  $P$-denaturation is accompanied  by a decrease of volume $\Delta V<0$  at high $P$ and an increase of volume $\Delta V>0$ at low $P$.
In particular, at $P=0.3(4\epsilon/v_0)$, corresponding to $\approx 500$~MPa, we find that
$\Delta V\approx -2.5 v_0$, hence $|P\Delta V|=0.75 (4\epsilon)\approx 17$~kJ/mol, very close to the typical reported value of 15~kJ/mol \cite{MeersmanChemSocRev2006}.

Therefore our coarse-grain model allows to understand how water
contributes to the temperature-- and pressure--denaturation of
proteins.  Accounting for stronger and more stable HBs in the
hydrophobic hydration shell with respect to the bulk and for a more
compressible hydrophobic hydration shell our model reproduces a close
stability region for proteins with the expected elliptic-like shape in
the $T-P$ plane, consistent with theory \cite{hawley}. 
We find that cold denaturation is energy-driven, while unfolding at
high pressures and negative pressures are density- and enthalpy-driven
by water, respectively.  

\section{Protein adsorption onto NPs: kinetics of the protein-corona formation}

\begin{figure}
  \includegraphics[width=\linewidth]{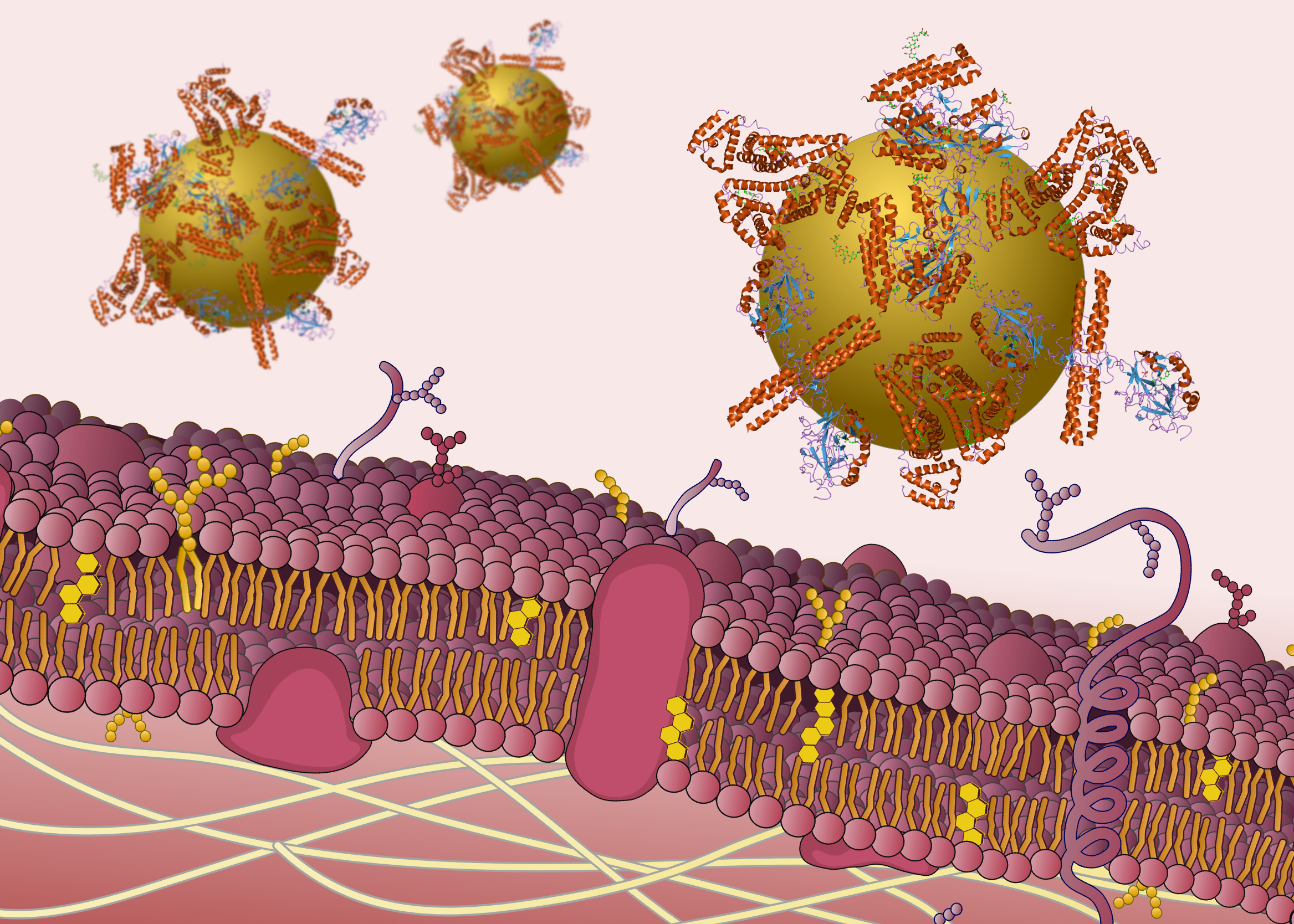}
  \caption{Pictorial representation of the NP-``Protein corona'' complex near 
  a cellular membrane}
  \label{fig:CoronaPicture}
\end{figure}

Nanoparticles (NPs) are small scale objects with dimensions in the range from 
  1 nm ($10^{-9}$ m) up to 100 nm. The main feature of these particles 
  rely on the fact that some of their properties differ completely from the bulk 
  material they are made of. Furthermore, their properties strongly
  depend on the particle size.  
  Due to their small size, the surface to volume ratio of NPs is much higher than 
  that of macroscopic objects.  This feature, combined with the high surface free 
  energy of NPs, confers NPs a high level of chemical reactivity.
  
  The NPs particular interactions with biological systems make them 
  very promising tools for 
  medical applications and could allow to simultaneously perform
  therapeutics and diagnostics ({\it theranostics}) \cite{Habash1999,Salvati2013,Ding2016}. 
In particular, experiments show that NPs are 
able to cross cellular barriers, including the strongest defense we
have in our body, the blood-brain-barrier. 
Therefore, the fact that NPs  interact directly with the biological
machinery \cite{DeSimone2006} represent an opportunity to deliver
drugs to specific targets hidden in the
  most inaccessible spots within the cells for  treating  illnesses
  that challenge us, such as      
  cancer or neurodegenerative diseases  \cite{Puntes2015}.

In the last years, industry has started to produce NPs at a large scale, with an 
increasing rate, due to their industrial, commercial and medical applications.
Thousands of commercial products, such as textiles, cosmetics or paints contain 
NPs, and nanomaterials are nowadays used in medical treatments, electronics, 
or food. In order to keep producing and using them at the industrial
scale safely,  many experimental studies \cite{Lindman2007}
of the nanotoxicology impact \cite{Dawson:2009kx}
of specific NPs have started to evaluate the hazard of exposing the environment,
living beings and humans to them \cite{Lundqvist:2008vn, Pratap09,
  RiveraGil2010,Corbo2016}.  

Yet, very little is known about the mechanisms regulating the interactions of NPs
with biological systems \cite{Lynch:2007kx}.
Acquiring such knowledge would allow us to predict if the interaction of such small-scale 
 materials with living organisms would potentially be dangerous
before even performing elaborated experiments.
 
 It has been confirmed that NP's absorb proteins and other biomolecules 
 from the environment forming a complex that is known as the 
``protein corona'' (Fig.\ref{fig:CoronaPicture}) 
\cite{Cedervall2007, Lynch2009, Walczyk2010,
   Casals2010, DellOrco2010, Milani2012}. Certain proteins are 
 also able to prevent the adsorption of other molecules, or to modify the 
 chemical properties of the NP surface, and even to determine the path and 
 final localization of the NP in a living organism \cite{Pitek2012,Salvati2013}.

Material surfaces exposed to biological environments are commonly modified 
by the adsorption of biomolecules, such as proteins and lipids, already 
present in solution. It has been shown that the cellular 
response to any material in a biological medium is mediated by the 
adsorbed biomolecular layer, rather than the bare material itself \cite{Lynch2009,Walczyk2010}. 
For this reason, the scientific community has focused its attention to the fact that NP interactions 
with living organisms must be also mediated by the adsorbed protein
layer \cite{Monopoli2012, Lundqvist2011}. It is now understood that
the biological identity of the NP is characterized by the distribution of 
the distinct adsorbed proteins on its surface \cite{Shapero2011,
  Salvati:2011ys, Monopoli2011, Monopoli:2011}. This collection of adsorbed 
biomolecules is in fact dynamic and evolves in time
\cite{Lynch2009,Walczyk2010,Milani2012,Casals2010, Lundqvist2011}.
The effective particle made by the protein-NP-corona is essential to 
understand how NPs interact with cells in biological media \cite{Monopoli2012}.
   Our aim is to understand and explain how the interactions between biological 
  macromolecules and NPs in solution leads to the formation of the effective 
  particle and to its evolution over   time scales that are
  relevant in the biological context.

In the following we address the establishment of a 
theoretical framework and the determination of the key features that
could give a simple, but complete, picture of protein-NP interactions.
We focus on parameters that describe the NP surface at the molecular level, such as curvature, 
charge or surface chemistry, that determine which kinds of proteins ends up binding from 
a complex biological fluid, e.g., blood plasma or cell media \cite{Monopoli2011}. 

The adoption of a multi-scale approach in this respect is particularly
suitable because it provides the theoretical framework to 
study a variety of aspects that occur at different length- and time-scales, ranging from the 
atomistic scale ($\sim 0.1$nm and $\sim 1$ns) to the mesoscale ($\sim
1\mathrm{\mu}$m and $\sim 100$s).
We, therefore, 
first  establish phenomenological models for protein-protein 
and  protein-NP interactions, in solvent \cite{Vilaseca2013}. 
Next, we validate these models with, at least, two types of NPs using available 
experimental data \cite{VilanovaPhDThesis}.
Finally we perform predictions, based on our theoretical models, and 
we test them by comparing with experimental data. In particular,
experimentally checking  the kinetics and the  protein 
corona composition allows us to assess the predictive power of the
models when we use only our knowledge on 
physico-chemical properties of the nanomaterial 
and the environmental conditions \cite{Vilanova2016}.

\section{ Methodology}

\subsection{Multi-scale modeling}

For the purpose of developing this research we make use of state-of-the-art 
techniques of computational and theoretical modeling. Thereby, we adopt a 
multi-scale approach to carry out this study. Multi-scale modeling is based on 
the idea that each specific problem is characterized by multiple scales of 
time and space. 
There is a need to consider every scale to understand the whole picture, given the 
relevance of different mechanisms occurring at each scale. This approach 
is particularly appropriate for the case of biological systems at the nanoscale, 
such as the interactions of NPs with macromolecules in presence of water. In this 
situation, the length scales range from angstroms to micrometers while the time 
scales span more than ten orders of magnitude. 

We separate our analysis into different levels of description, based on the 
observation that each length scale has an associated time scale. By these means, 
each level of description is related to a range of length and time scales that 
partially overlap with the consecutive levels. To this end, each level is 
characterized by specific phenomena, which can be useful to understand features of 
other, maybe more complex, phenomena at higher levels of description.
In our case, we consider the following levels of description: (i) the molecular 
and macromolecular level, ranging from 1 $\AA$ to 100 nm and from 10 fs to 
1 $\mu\mr{s}$ , and (ii) the mesoscale level, ranging from 1 nm to 1 $\mu\mr{m}$ 
and from 10 ns to 1 s. The next step is to consider, for example, the problem 
of NP aggregation or the interaction of NPs with cellular membranes, that would 
span a range of length scales up to 100 $\mu\mr{m}$ and time-scales up to hours.

\subsection{Theoretical framework}
We make use of suitable computational and theoretical techniques 
to approach each level of description. The smaller scales are characterized by 
the explicit consideration of water molecules in the solvent, as mediators of 
the processes that occur at the macromolecular scale. 
However, dealing with explicit solvent models hampers the efficiency of 
computation in a dramatic way. This is why we adopt the coarse-grain model 
of water introduced in the previous sections \cite{StokelyPNAS2010,MazzaPNAS2011,BiancoSR2014} to describe
the macromolecule-solvent interactions \cite{BiancoPRL2015}. This is  
a convenient strategy that allows for wider and more efficient options to explore 
these systems. The model can be efficiently simulated using the 
Monte Carlo (MC) method
\cite{Kumar2008c,Mazza2009,StokelyPNAS2010,MazzaPNAS2011} and
can be also treated analytically, allowing us to reach  a  
deeper understanding of the fundamental mechanisms
\cite{Franzese2002,Franzese2002a,FS2002}.  

At the larger scales, where we focus our attention on the interaction of NPs with 
solutions containing a large number of proteins, we use an implicit solvent 
approach. 
In this sense, we take into account the effects of the solvent by my means of 
modified dynamics and effective interactions. Therefore, we also consider the 
proteins and the NPs as coarse-grain objects. We adopt a Molecular Dynamics (MD) 
simulation scheme using Langevin dynamics (LD), that allows for the determination 
of the kinetic properties of the system. Moreover, we introduce a framework that 
describes protein-protein interactions via effective potentials, and we make use 
of the well established DLVO theory for colloidal dispersions in solution to 
describe the protein-NP interactions \cite{Vilanova2016}. 

Finally, we develop a Non-Langmuir Dynamical Rate Equation (NLDRE) model which is 
 phenomenological analytical theory to describe the protein adsorption kinetics \cite{Vilanova2016}. 
 This theory provides us tools to extrapolate the simulation results of 
protein adsorption obtained with MD, to much longer time-scales. With
this approach
we are able to predict kinetic processes over experimentally relevant
time-scales of the order of several minutes, far beyond the limits of
the  computational capacities \cite{Vilanova2016}. 

\subsection{Experimental validation}

The collaborative work of theorist and experimentalists is a key factor 
for the successful development of theoretical models with predictive
capability.
The experimental data is fundamental for the parametrization of the 
theoretical models, yet it is crucial for the validation and verification 
process of the theoretical predictions.
To this goal we focus on a simplified version of a multicomponent 
protein solution as it would be the human blood plasma. In particular,
we adopt a ``model plasma'' containing only three proteins that are 
representative of the extremely large number of protein forming the
human blood \cite{Vilanova2016}. This step is essential 
to design experiments suitable for comparison with simulations.

We follow a workflow where we can first  measure in a controlled
experimental setup all the data that are relevant for defining the
phenomenological parameters of the theoretical model.
In this way, the preliminary experimental results serve to calibrate 
the theoretical model.

Next we perform our simulations and analytic prediction based on these
phenomenological input parameters. We make our calculation under
conditions that can be reproduced in experiments, in simple cases with
NPs in bicomponent protein solutions, for a direct comparison.
Then we design a set of simple experiments to validate the 
simulation results obtained under identical conditions.

Finally, to test the predictive power of the theoretical tools,  
we consider more complex situations, such as NPs in three-component
solutions made of proteins that are competing for the NP surface.
Under these conditions the experiments reveal a \emph{memory effect}
of the protein corona, not predicted by our initial model. 
As a consequence, we modify the model in a way that allows us to
reproduce the effect. Although this result exposes a limit of our
initial approach, it also led us to propose a mechanism to rationalize
the memory effect. Specifically, we show that to account for the
memory effect
it is necessary to model the self-assembly of the proteins on the NPs
including irreversible structural changes \cite{Vilanova2016}.

\subsection{Simulation details and results}

\begin{figure}
  \includegraphics[width=\linewidth]{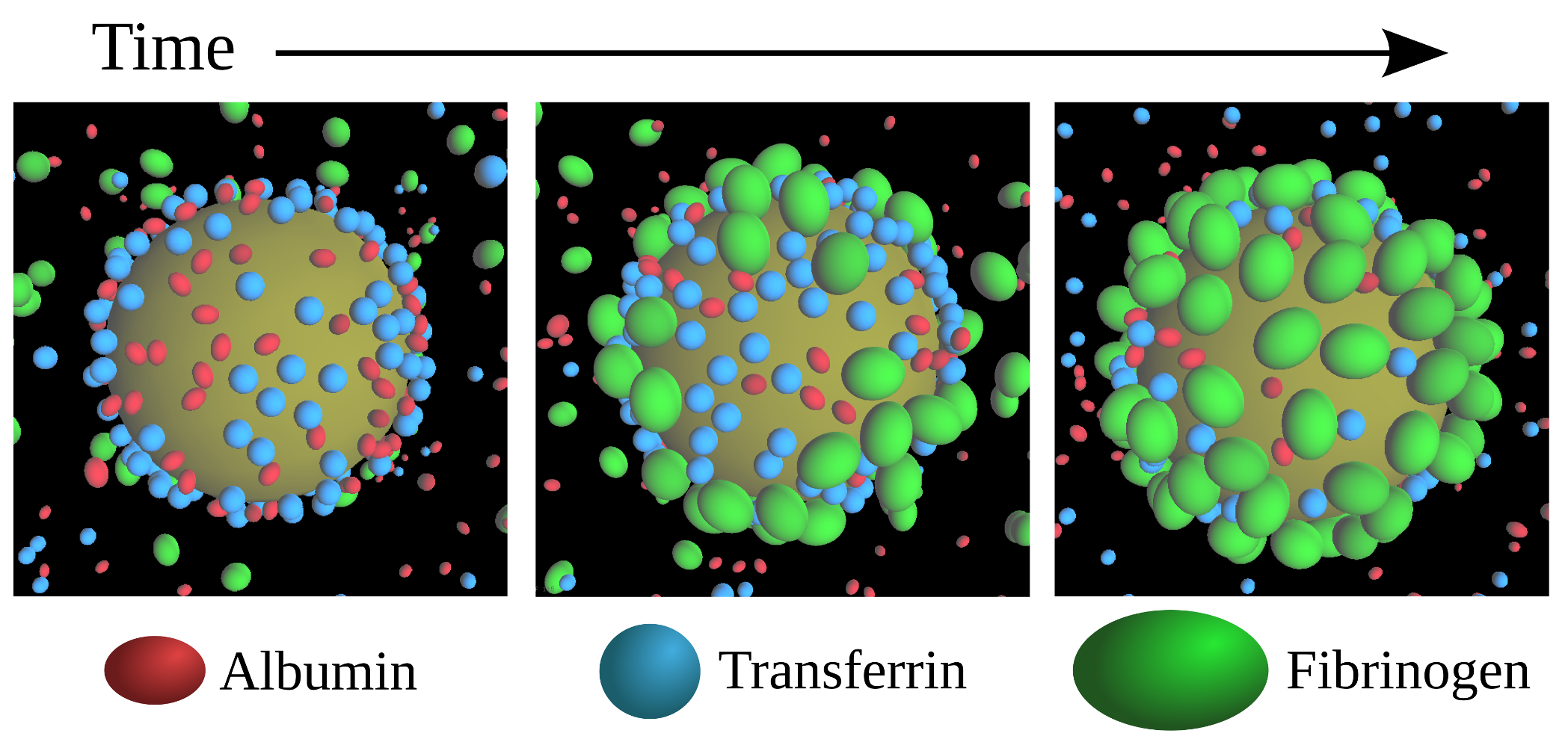}
  \caption{Simulation snapshots at three different times for a system containing 
  one silica NP of 100nm diameter in a solution containing Human Serum Albumin (HSA), 
  Transferrin (Tf) and Fibrinogen (Fib).
  The introduction of each protein is done sequencially 
  in three steps: HSA first, Tf second and Fib third.}
  \label{fig:SimSnapshots}
\end{figure}

We develop a high-performance suite of simulation codes able to run on Graphical 
Processing Units (GPU's) \cite{Bubbles2015}. We use a molecular dynamics (MD) approach to 
simulate the interactions of NPs with protein solutions. 
We consider a simulation box with periodic 
boundary conditions containing one single NP fixed to the center of the box.
We separate the simulation box into two regions. The inner region contains 
the NP, while we use the external region as a reservoir to control the protein 
concentration inside the inner region \cite{Vilanova2016}. 
We treat the solvent in an implicit way, by introducing a coarse-grain protein-protein 
and protein-NP interaction model using continuous shouldered
potentials \cite{Vilaseca2013, Fr07a, Vilaseca2011}.
We also adopt a Langevin dynamics (LD) integration scheme to take into account 
the collisions with the solvent molecules for the correct diffusion of the proteins.

\begin{figure}
  \subfloat[]{\includegraphics[width=\linewidth]{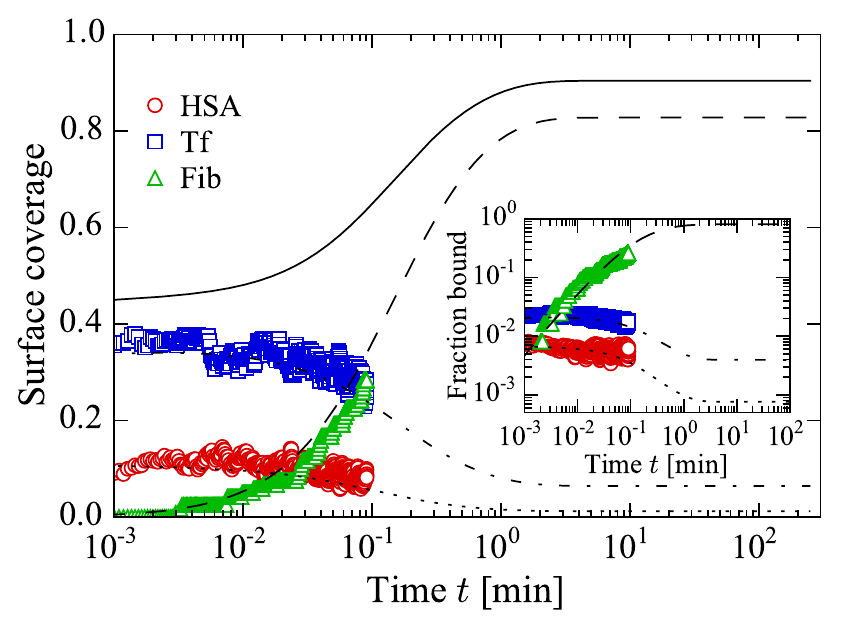}}\\
  \subfloat[]{\includegraphics[width=\linewidth]{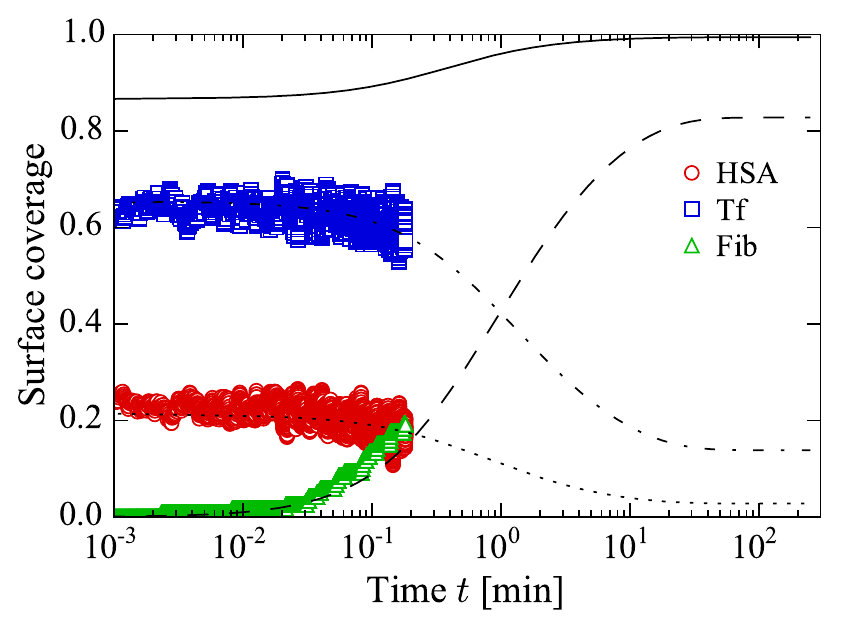}}
  \caption{Simulation results (symbols) and analytical extrapolation (lines) using 
  NLDRE of the protein corona kinetics in a system containing 100nm silica NPs at 0.1 mg/ml in 
  a protein solution containing HSA (red circles), Tf (blue squares) and Fib (green triangles) 
  at different concentrations. The NPs are initially incubated in a solution 
  containing HSA and Tf, with the subsequent introduction of Fib at 0.005 mg/ml. 
  We also show the analytical extrapolation using the NLDRE theory to the 
  simulation data for each protein (dashed lines), and the total surface coverage (solid lines).
  a) NP surface coverage after 
  incubation in a solution of HSA at 0.07 mg/ml and Tf at 0.07 mg/ml, inset: Fraction bound 
  of proteins shown for comparison with the surface coverage. 
  b) NP surface 
  coverage after incubation in a solution of HSA at 3.5 mg/ml and Tf at 3.5 mg/ml.  }
  \label{fig:HSAwTfwFib}
\end{figure}

We follow a sequential protocol to insert different kinds of proteins at selected 
times during a simulation. For example, in Fig.\ref{fig:SimSnapshots} we show 
three snapshots of a simulation containing a 100nm diameter silica (SiO$_{2}$) NP at 0.1 mg/ml
in a protein solution of Human Serum Albumin (HSA), Transferrin (Tf) and Fibrigonen (Fib). 
The first snapshot shows the equilibrium state after introducing HSA at 0.07 mg/ml 
and Tf at 0.07 mg/ml to the system, at the same moment that we introduce Fib at 0.005 mg/ml. 
The second snapshot is a transient state, where Fib begins adsorbing to the NP surface.
In the third snapshot we show how the high affinity of Fib allows it to displace 
HSA and Tf proteins from the surface, while the system has not yet reached 
equilibrium because this is a kinetically slow process
 (Fig.\ref{fig:HSAwTfwFib}a).

\captionsetup[subfigure]{position=top, labelfont=bf,textfont=normalfont,singlelinecheck=off,justification=raggedright}

In Fig.\ref{fig:HSAwTfwFib} we show the effect of increasing the concentration 
of the proteins during the incubation step to the adsorption kinetics of the third
protein. In panel a) the NP is incubated in a solution of HSA and Tf at a low 
concentration of 0.07 mg/ml each, and we introduce Fib at 0.005 mg/ml which is 
a protein with a much higher surface affinity. We make use of our NLDRE to fit 
the parameters to the simulation data and we extrapolate the adsorption kinetics 
to much longer time-scales. In panel b) we show the adsorption kinetics of a system 
where we initially incubate the NP in a solution of HSA and Tf at a much higher 
concentration of 3.5 mg/ml each. Using the analytical extrapolation to the 
simulation data we find that the Fib adsorption kinetics are at least 1 order of 
magnitude slower compared to the previous situation. 

\section{Conclusions}

 We combine theory with experiments to introduce a new framework consisting in 
 experiment design supported by computational simulations and theoretical models 
 as a methodology to obtain reliable predictive results for protein
 folding and  protein-NP self-assembly.

We establish a multi-scale modeling approach to deal with the different phenomenologies 
characteristic to each level of description.
Based on atomistic Molecular Dynamics simulations over nanoseconds up to microseconds
time-scales, we define a coarse-grain water model that allows us to
simulate hydrated systems \cite{Vilanova2011}
at length- and time-scales that are relevant
for biological processes. This model allows us to study by Monte Carlo
simulations protein structural changes and folding-unfolding events
\cite{BiancoPRL2015}. We recently extended this approach to study
protein design \cite{BiancoPRX2017}.

With this in mind, we  introduce 
a set of computational tools (BUBBLES \cite{Bubbles2015}) that allows
us to study phenomenological protein-protein and protein-NP 
interactions \cite{Vilaseca2013}
with Langevin Dynamics simulations up to the time scale
of seconds. Next 
we introduce a phenomenological theory 
based on rate equation (NLDRE) with which we can  extrapolate our simulation results 
to time scales of hours, allowing us direct comparison with the
experimental results. 
 
In particular, we first study the protein corona formation for a simple system of polystyrene NPs in 
a solution containing only a single kind of proteins (Transferrin). We introduce 
many-body interactions to explain the multilayer adsorption 
mechanism, the protein corona kinetics, and the soft-corona/hard-corona 
characterization \cite{VilanovaPhDThesis}.

Next, we study more complicated systems by introducing NPs 
in solutions made of multiple types of proteins, choosing a set of
proteins that compete for the NP surface.
This allows for the emergence of competitive protein adsorption and
assembly on top of the NP, a rich and 
complex playground that we exploit to  discover and understand new and 
unexpected features \cite{Vilanova2016}.
\newline

\section*{Acknowledgement}
We are thankful to 
M. Bernabei, C. Calero, L. E. Coronas, F. Leoni, N. Pag\`es, and A. Zantop
for helpful discussions.
O.V. and G.F. acknowledge the support of Spanish MINECO grant
FIS2012-31025 and FIS2015-66879-C2-2-P.
I. C.  acknowledges the support from the Austrian Science Fund
(FWF) Grant P 26253-N27. V.B. acknowledges the support of the FWF
Grant 2150-N36 and P 26253-N27.

\end{document}